\documentclass[12pt]{iopart}
\usepackage{iopams}  

\usepackage{graphicx}

\usepackage{color}
\usepackage{ulem}
\usepackage{rotating}
\usepackage{sidecap}

\newcommand{\eps}[0]{\varepsilon}	

\renewcommand{\vec}{\mathbf}

\renewcommand{\vec}[1]{\mathbf{#1}}

\begin{document}

% \keywords{plasma crystals, streaming effects, Yukawa balls, string formation, dynamical screening}

\title[Ion-Streaming Induced Order Transition]{Ion-Streaming Induced Order Transition in 3D Dust Clusters}

\author{Patrick Ludwig, Hanno K\"ahlert, and Michael Bonitz }

\address{Institut f\"ur Theoretische Physik und Astrophysik, Christian-Albrechts-Universit\"at zu Kiel,
 Leibnizstra{\ss}e 15, 24098 Kiel, Germany}

\ead{ludwig@theo-physik.uni-kiel.de}

\begin{abstract}
Dust Dynamics Simulations utilizing a dynamical screening approach are performed to study the effect of ion-streaming on the self-organized structures in a three-dimensional spherically confined complex (dusty) plasma.
Varying the Mach number $M$ -- the ratio of ion drift velocity to the sound velocity, the simulations reproduce the experimentally observed cluster configurations in the two limiting cases: 
at $M=0$ strongly correlated crystalline structures consisting of nested spherical shells (Yukawa balls) and, for $M\geq1$, flow-aligned dust chains, respectively. In addition, our simulations  reveal a discontinuous transition between these two limits. It is found that already a moderate ion drift velocity ($M\approx0.1$) destabilizes the highly ordered Yukawa balls  and initiates an abrupt melting transition. The critical value of $M$ is found to be independent of the cluster size.
\end{abstract}

\maketitle

\section{Introduction}
Collective dynamical effects such as wakefields are omnipresent in nature. One of the most common phenomena is the V-shaped `Kelvin wake pattern' created by watercrafts moving through water. In a plasma, similar wake pattern give rise to 
an effective attractive interaction between particles that otherwise mutually repel each other. In recent years, plasma wake fields which are driven by an ultrashort laser pulse are used for the acceleration of electron beams where it is expected that rates that are more than a thousand times higher than those achieved in conventional accelerators may be achievable~\cite{laserwake1,laserwake3}. 
In an entirely different context -- conventional superconductors -- similar attractive forces are created by the dynamically polarised background medium giving rise to the Cooper pairing of electrons. 

An excellent test system for many-particle correlations and structure formation with pronounced streaming and wake effects
are complex (dusty) plasmas \cite{rev2010}.
In the plasma bulk of an radio-frequency (RF) discharge, where the plasma flow is negligible, spherical 3D plasma crystals (Yukawa balls) with a nested shell structure have been created~\cite{arp}. A completely different type of dust structure is observed in the regime of supersonic ion flow, $M > 1$.
In the plasma sheath region, where strong electric fields are present, the grains form vertically aligned `dust molecules' and, with the help of a confining potential, long particle strings can be created, e.g. \cite{taka,kroll,killer}. The vertical alignment of particles is not explainable with a purely repulsive grain interaction. Therefore, it was recognized early that the highly charged grains have a focusing effect on the streaming ions and a positive ion space charge is accumulated in the wake below the grains~\cite{vlad}. Experimentally, the existence of an attractive wake potential and a resulting non-reciprocal grain interaction was demonstrated by means of laser manipulation experiments on particle strings~\cite{taka} and dust molecules~\cite{kroll,melzer2001}.

% M unbekannt!!!!
While the two limiting cases (Yukawa balls and flow-aligned dust strings) have been explored experimentally and theoretically in great detail, see e.g. \cite{rev2010,mebu2010} and references therein, a key question has-- to our knowledge --not been addressed so far: how does the structural order transition proceed between both limits when the ion flow is gradually increased and wake effects increasingly take charge?
Experimentally, this question may be hard to answer since the ion-streaming velocity is neither directly accessible nor separately tunable without changing other plasma parameters at the same time. Therefore, computer simulations may help to answer these questions. However, a self-consistent numerical simulation of a nonideal multi-component plasma in non-equilibrium is highly challenging due to the drastic differences in the time scales of the plasma constituents: electrons, positively charged ions, neutral atoms, and negatively charged `dust' grains of micrometer size.
Utilizing the `Dynamical Screening Approach' (DSA) this multiscale problem can, however, be effectively reduced by an approximate (statistical) description of the lighter plasma constituents~\cite{vlad,lampe2000,joyce2001,lampe2005,jenko2005, wakestruct}. The DSA allows for an accurate description of essential equilibrium and non-equilibrium plasma properties including plasma streaming induced wakefield oscillations, ion and electron thermal effects as well as collisional and Landau damping. On this basis, the dust dynamics will be studied by first principle Langevin Dynamics simulations, which allow for a full dynamic treatment of the dust pair interaction. 

This article is outlined as follows: 
In section \ref{grainpot} the DSA is employed to incorporate the dynamical screening properties of the streaming plasma background into an effective
electrostatic potential of a single grain. 
In section \ref{dds} we then use these results for many-particle dust dynamical simulations which reveal the collective particle behavior for a wide range of ion-streaming velocities. A final discussion of the ion-streaming induced order transition follows in section \ref{conclusion}.

\section{Dynamically Screened Grain Potential} \label{grainpot}
Collective dynamical effects of the highly charged, massive dust particles are mainly determined by their electrostatic interactions.
For the computation of the electrostatic potential of a single grain we consider plasma parameters that match the typical experimental conditions:
plasma density $n_e=n_i=10^{8}/\,\mathrm{cm}^{3}$,
electron temperature $T_e=2.5\,\mathrm{eV}$, ion temperature $T_i=0.03\,\mathrm{eV}$. 
A moderate Argon gas pressure of $p=15\,\mathrm{Pa}$  corresponds to an ion-neutral collision frequency of $\nu_{in}=0.2\,\omega_{\mathrm{p}_i}$ \cite{nuin}, where $\omega_{\mathrm{p}_i}=\sqrt{n_i q_i^2/(\eps_0 m_i)}=2.1\cdot 10^6\,\mathrm{Hz}$ is the plasma frequency of singly charged Argon ions of mass $m_i=6.634\cdot 10^{-26}\,\mathrm{kg}$.

\begin{figure}
\includegraphics[width=0.33\textwidth]{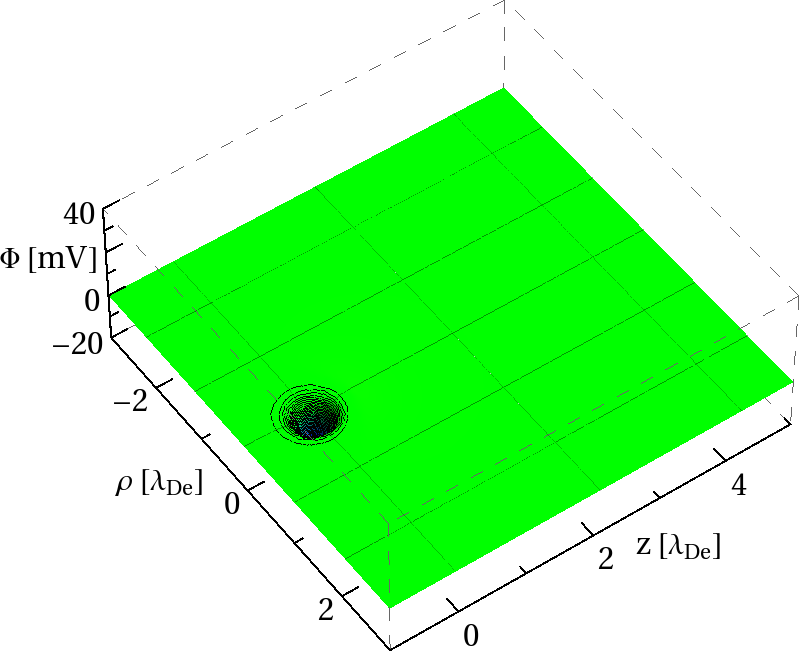}~\hspace{-15mm}\begin{scriptsize}a)~M=0.01\end{scriptsize}
% \hfil
\includegraphics[width=0.33\textwidth]{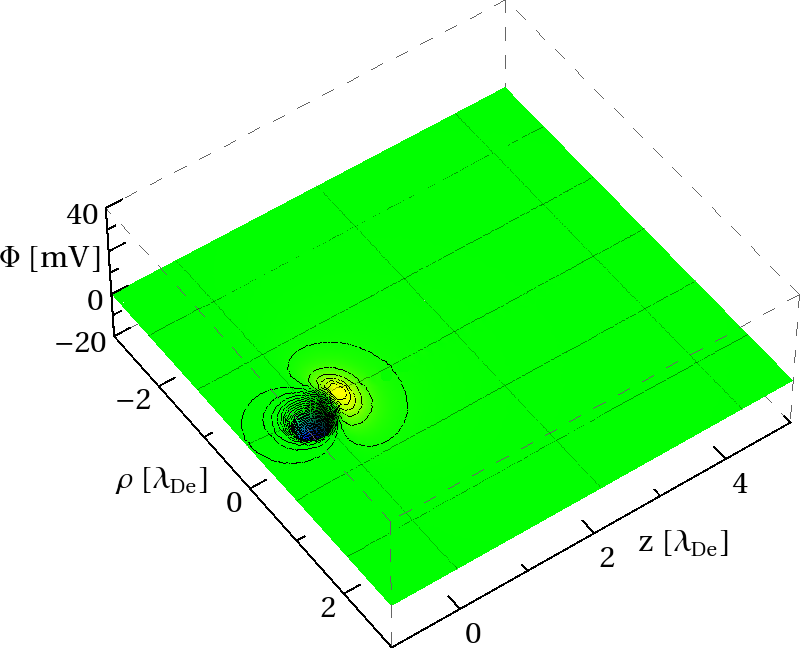}~\hspace{-15mm}\begin{scriptsize}b)~M=0.08\end{scriptsize}
\includegraphics[width=0.33\textwidth]{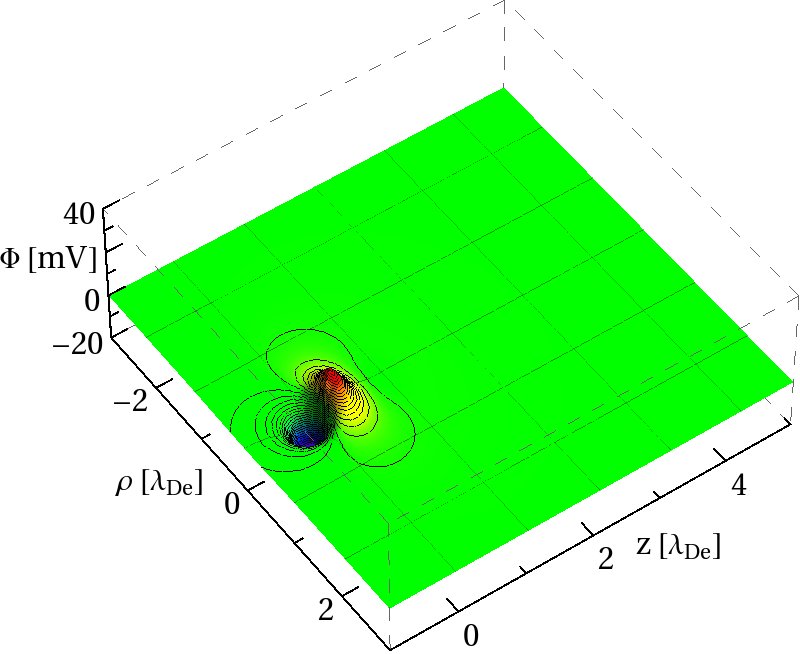}~\hspace{-15mm}\begin{scriptsize}c)~M=0.16\end{scriptsize}
\includegraphics[width=0.33\textwidth]{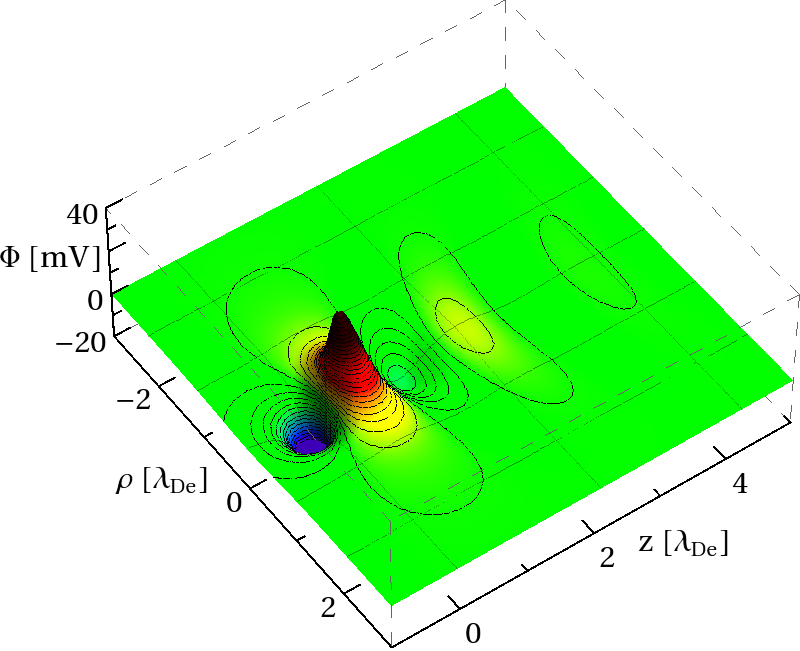}~\hspace{-15mm}\begin{scriptsize}d)~M=0.33\end{scriptsize}
\includegraphics[width=0.33\textwidth]{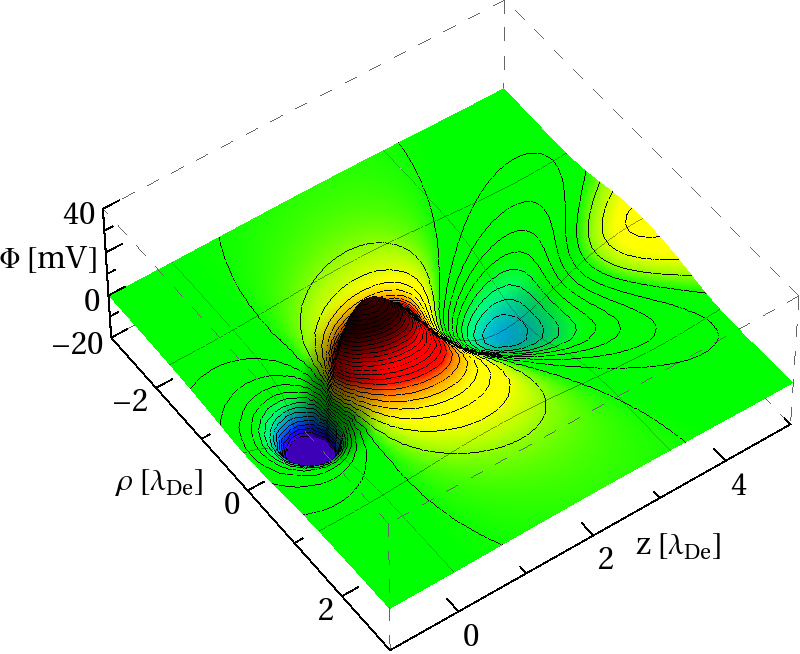}~\hspace{-15mm}\begin{scriptsize}e)~M=0.66\end{scriptsize}
\includegraphics[width=0.33\textwidth]{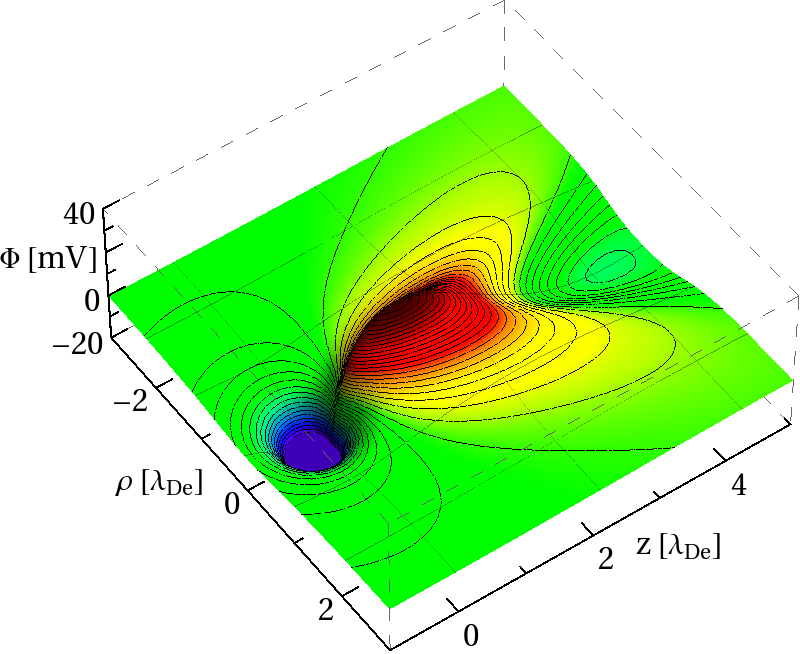}~\hspace{-15mm}\begin{scriptsize}f)~M=1.00\end{scriptsize}
% \hfil
% \hspace{-42mm}
 \includegraphics[width=0.995\textwidth]{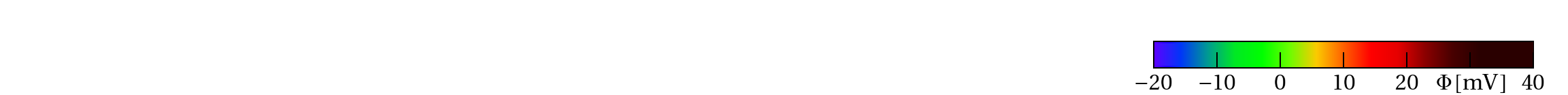}
\caption{(Color online) Dynamically screened dust (wake) potential $\Phi(\vec r)=\Phi(\rho,z)$ as obtained from Eq. (\ref{potential}), $\lambda_\mathrm{D_e}=0.128\,\mathrm{mm}$. The grain is located at $\{\rho,z\}=\{0,0\}$. The ions stream along the z-direction with $M=0.01,0.08,0.16,0.33,0.66,$ and $1.0$,  respectively, giving rise to a wake structure behind the grain. The values of the potential along the $z$-axis are shown also on Fig.~\ref{fig:wakesideplot}.}
\label{fig:grainpot}
\end{figure}

Assuming a weak (linear) response of the plasma to the presence of the dust grain, the dynamically screened grain potential in a streaming plasma is obtained in real space by means of a  3D Fourier transformation 
\begin{equation} \label{potential}
\Phi(\vec r)   = \int\!\mathrm{d}^3{\vec{k}} \frac{q_d}{2 \pi^2 k^2 \epsilon(\vec k, \vec k \cdot \vec u_i)} e^{i \vec k \cdot \vec r} \quad,
\end{equation}
where $\vec r$ denotes the distance from the grain's center of mass~\cite{alexandrov}.
For the calculation of the wake potential we use a grain charge of $q_{\mathrm{d}}=-10^5\,e_0$, but the potential can be rescaled linearly to any other charge value, see section \ref{dds}. 
The ambient plasma is represented by a longitudinal dielectric function which includes a directed ion flow $\vec{u}_i$ and Bhatnagar-Gross-Krook (BGK)-type ion-neutral collisions~\cite{jenko2005}
\begin{equation} \label{dk}
\epsilon(\vec{k},\omega)= 1 + \frac{1}{k^2 \lambda_{\mathrm{D}_e}^2} + \frac{1}{k^2 \lambda_{\mathrm{D}_i}^2} \left[ \frac{1+\zeta_i Z(\zeta_i)} {1+\frac{i\nu_{in}}{\sqrt{2} k v_{T_i}}Z(\zeta_i)} \right] \quad.
\end{equation}
Here, we employ the substitution
$\zeta_i=(\vec{k}\cdot\vec{u}_i+i\nu_{in})/(\sqrt{2} k v_{T_i})$ and the plasma dispersion function
$ Z(z)= \pi^{-\frac{1}{2}}\int_{-\infty}^\infty \mathrm{d}t \frac{exp(-t^2)}{t-z}$~\cite{wakestruct}. Further, we introduced the 
Debye length of electrons $\lambda_{\mathrm{D}_e}=1175.41\,\mathrm{\mu m}$ and ions $\lambda_{\mathrm{D}_i}=128.76\,\mathrm{\mu m}$, respectively, which are
 defined as $\lambda_{\mathrm{D}_\alpha}=(\eps_0 k_{\mathrm{B}} T_\alpha/n_\alpha q_\alpha^2)^{1/2}$. Note that screening by the electrons is purely static, since their thermal speed exceeds their field-induced drift. 
The ion thermal speed $v_{\mathrm{T}_i}\equiv\sqrt{k_{\mathrm{B}} T_{i}/m_i}=270\,\mathrm{m/s}$ is well below the 
ion sound (Bohm) speed $c_s\equiv\sqrt{{k_{\mathrm{B}} T_e}/{m_i}}=2460\,\mathrm{m/s}$.  The ion streaming velocity is expressed in terms of the Mach number $M \equiv |{\vec u}_i|/c_s$.

\begin{figure}[t]
\includegraphics[width=0.8\textwidth]{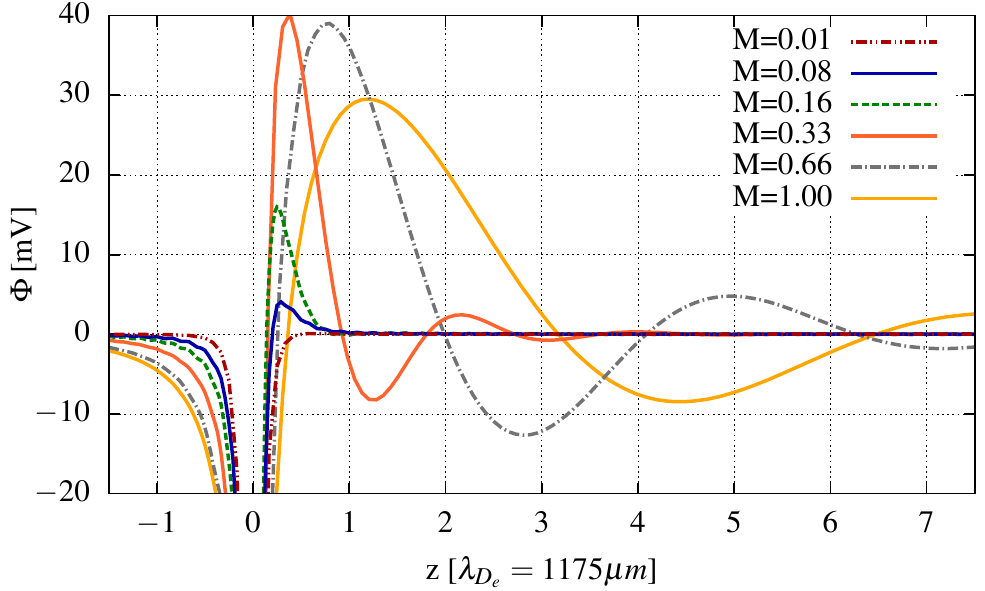}
\caption{(Color online) Electrostatic grain potential $\Phi(r,z)$ in streaming direction $z$ at the radial position $\rho=x=y=0$ [cut through Fig.~\ref{fig:grainpot}].}
\label{fig:wakesideplot}
\end{figure}

In Figs.~\ref{fig:grainpot} and \ref{fig:wakesideplot}, we present the dynamically screened Coulomb potential for a single grain in the relevant range of ion flow velocities $\vec u_i$.
When ion drift and collisions are negligible, that is $M \rightarrow 0$ and $\nu_{in} \rightarrow 0$, Eq.~(\ref{dk})
reduces to static electron and ion shielding of the Coulomb potential and the grain potential becomes the isotropic Debye-H\"uckel (or Yukawa) potential
$\Phi_{\rm Yuk}(r)=q_d \, e^{-r/\lambda_{\mathrm{D}}}/(4\pi \eps_0 r)$, where $\lambda_{\mathrm{D}}  \equiv (\lambda_{\mathrm{D}_e}^{-2}+\lambda_{\mathrm{D}_i}^{-2})^{-1/2}=128.00\,\mathrm{\mu m}$  is the combined Debye length of electrons and ions. Fig.~\ref{fig:grainpot}a) shows a 3D plot of the electrostatic potential $\Phi(\vec r)=\Phi(r,z)$ for $M \approx 0$ and finite pressure.

A stationary flow of ions along the z-direction leads, however, to strong deviations from the Yukawa potential by giving rise to an anisotropic, oscillating wake structure behind each grain, as shown in Fig.~\ref{fig:grainpot}b)-f). With increasing $M$ the amplitude of the ion focus increases and reaches its maximum for $M \approx 0.5$, where $\Phi_{\rm max}=43\,\mathrm{mV}$. This means that essential wake effects are already present well below the supersonic regime. A consideration of an increasing dust charge $q_d$ with $M$ does not change the wakefield significantly, see OML results in Ref.~\cite{wakestruct}. In Fig.~\ref{fig:grainpot}d),  the first minimum and the second maximum of the potential become clearly visible in the wake structure. As a general trend, it is found that with increasing $M$ the primary ion focus and all higher extrema shift linearly away from the grain. In situations, where the streaming velocity is close to or exceeds the sound speed, the ion focus flattens out and a more and more stretched Mach cone forms. 
More details on the numerical implementation, the wake structure, and a critical assessment of the DSA will be published separately in Ref.~\cite{wakestruct}

\section{Dust Dynamics Simulations} \label{dds}
The dust grains are embedded in a partially ionized plasma and exposed to collisions with the neutral gas atoms.
Instead of treating the dust-neutral collisions explicitly, the Langevin scheme provides a stochastic approach, which involves the neutral gas as a large heat bath for the dust particles. The equation of motion for the k-th dust grain reads
\begin{equation} \label{eom}
m_d \ddot {\vec r}_k=- \nabla V_k^{\mathrm{eff}}(\vec r,t) - \omega_0^2  m_d {\vec r}_k  - \nu_{dn} m_d  \dot {\vec r}_k + \vec f_k(t) \quad.
\end{equation}
Since Maxwell's equations are linear, the total dynamic potential acting on the k-th grain is just the superposition of the electrostatic (wake) potentials created by all other $(N_d-1)$ grains
\begin{equation}
V_k^{\mathrm{eff}}(\vec r,t)=\sum_{l\neq k}^{N_d} q_d \Phi_l(\vec r,t) \quad.
\end{equation}
A screening independent parabolic confinement is explicitly included in Eq.~(\ref{eom}). It mimics the combined effect of gravity, thermophoresis, and electric fields which in the experiments with Yukawa balls are used to achieve an approximately isotropic harmonic confinement for the dust grains.\cite{dusttrap} 
The friction term emulates a damping mechanism and a stochastic force, $\vec f_n(t)$,  accounts for random heating effects.
The random force and the (neutral gas) friction term are related by the fluctuation-dissipation theorem
\begin{equation}
\langle{\vec f_k^\alpha(t)}{\vec f_l^\beta(t')}\rangle=2 m_d \nu_{dn} k_B T_n \delta_{kl} \delta_{\alpha\beta}\delta(t-t') \quad,
\end{equation}
where $\alpha,\beta \in \{x,y,z\}$ and $k,l \in \{1,\ldots,N_d\}$.
The standard Langevin algorithm employed here is extensively described in Ref.~\cite{ott}. 
Connecting Langevin dynamics with the dynamical screening approach, the DDS scheme allows for inclusion of finite temperature effects as well as a full dynamical treatment of all dust-dust interactions on the basis of the effective wake potential Eq.~(\ref{potential})~\cite{lampe2005,QDSA}.
Besides the plasma parameters used for the computation of the grain potential, additionally the mass $m_d=9.1\cdot 10^{-14}\,\mathrm{kg}$, charge $q_{\mathrm{d}}=-6000\,e_0$, and radius $R_d=2.43\,\mathrm{\mu m}$ of $N_d$ identical grains as well as the trap frequency $\omega_0=7.0\,\mathrm{Hz}$ enter the simulation as input parameters.
A dust-neutral collision frequency of $ \nu_{dn}=5.3 \cdot \frac{R_d^2  \, p}{m_d} \,\sqrt{m_n/k_B T_n}=19.1\,\mathrm{Hz}$ drives the dust temperature towards the neutral gas temperature $T_n=T_i$.

\begin{SCfigure}[2][t]
\includegraphics[width=.52\textwidth]{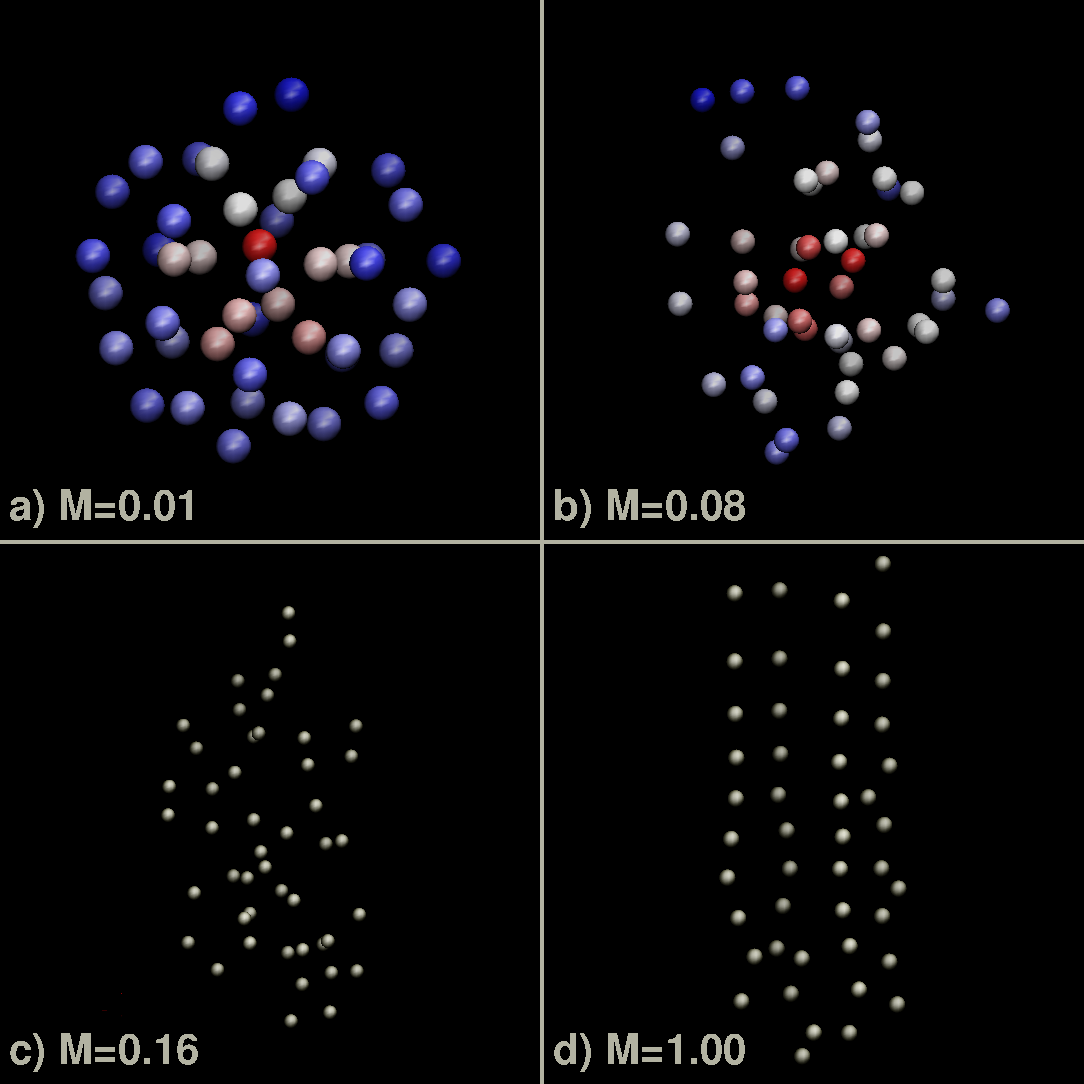}
\caption{
(Color online) Simulation snapshots of the (stationary) state of $N_d=46$ particles and the ion-streaming velocities $M=0.01, 0.08, 0.16,$ and $1.00$.  
The plasma streams from top to bottom. The color code for $M=0.01$ and $0.08$ denotes the particle's radial distance from the trap center. 
Approximate cluster size (in z-direction): a) $2.2\,\mathrm{mm}$, b)~$3.0\,\mathrm{mm}$, c)~$6.3\,\mathrm{mm}$, and d) $7.0\,\mathrm{mm}$.
}
\label{fig:dds}
\end{SCfigure}

In the following we will investigate the effect of the only remaining free parameter in the considered model system:
the relative streaming velocity of the ions with respect to the dust in terms of the Mach number $M$.
In the limiting case of no ion flow, i.e. $M=0$, the grain potential is of Yukawa-type. Starting from a random initial state, the grains equilibrate during the simulation due to frictional drag against the stationary neutral gas. Finally, the sub-system of dust grains reaches a strongly coupled state, where the dust forms a spherical Yukawa ball with a nested shell structure.
Fig.~\ref{fig:dds}a) shows the cluster with $N_d=46$ grains which exhibits a meta stable $(1,12,33)$ shell configuration for $M=0$. 
The highly symmetric icosahedral $12$ particle configuration in the inner shell is known to be particularly stable and was found to be energetically most favorable
for isotropically screened Coulomb particles in a parabolic confinement.\cite{baum}.
Earlier simulations in the frame of the one-component plasma model with a statically screened Coulomb potential $\Phi_{\rm Yuk}(r)$ have proven to yield very good agreement with the experiments in view of the structure~\cite{prl2006,springer}, dynamics~\cite{rev2010}, and melting behavior~\cite{schella} of finite dust crystals. 

Next we consider the case of supersonic ion-streaming. When the ions are streaming along z-direction from top to bottom,
they are deflected by the highly charged dust grains and induce an ion focus directly below the grain.
Other grains downstream are attracted to this positive space charge. As shown in Fig.~\ref{fig:dds}d), the ion wakefield attraction leads in the case of $N_d=46$ to the formation of four vertical particle chains. %  or, in the case of elongated clusters, to ther vertical alignment of flat rings. 
Because of the asymmetric nature of the (non-conservative) wake potential along the z-axis, the interaction between an upstream and a downstream grain is non-reciprocal~\cite{lampe2000,lampe2005}.
Kinetic energy of the streaming ions is coupled into the non-Hamiltonian dust system, which leads to self-excited vibrations and eventually to the instability of long chains~\cite{lampe2005,schweigert, ivth2011}. This effect is indicated in Fig.~\ref{fig:dds}d), where the dust particles in the upper part of the particle strings are highly ordered, but in the lower part grain dislocations due to a downward increasing, streaming-induced dust temperature are observed.

\begin{figure}
\begin{minipage}{0.48\linewidth}
\hspace{-2mm}
\includegraphics[width=1.\linewidth]{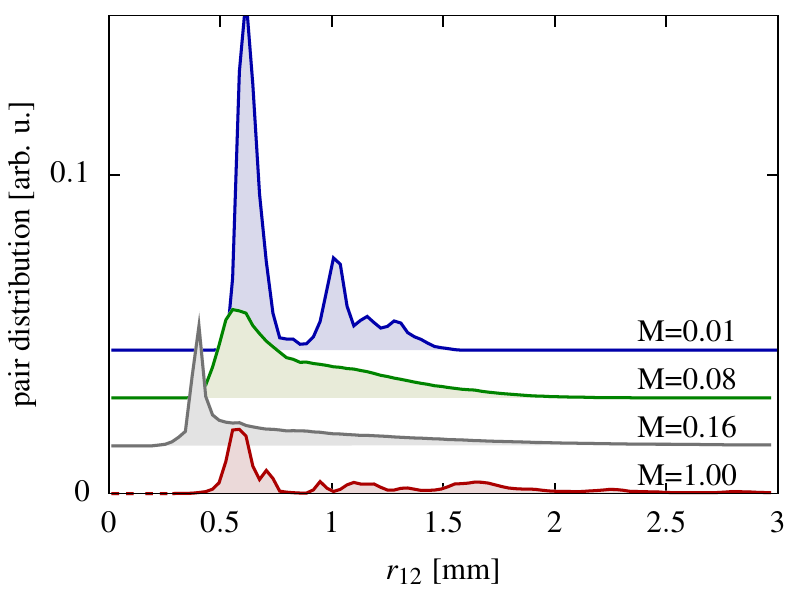}
\caption{(Color online) Pair distributions for $N=18$ grains at the streaming velocities $M=0.01, 0.08, 0.16,$ and $1.00$.}
\label{fig:g12}
\end{minipage}
\hfil
\begin{minipage}{0.48\linewidth}
\hspace{-2mm}
\includegraphics[width=\linewidth]{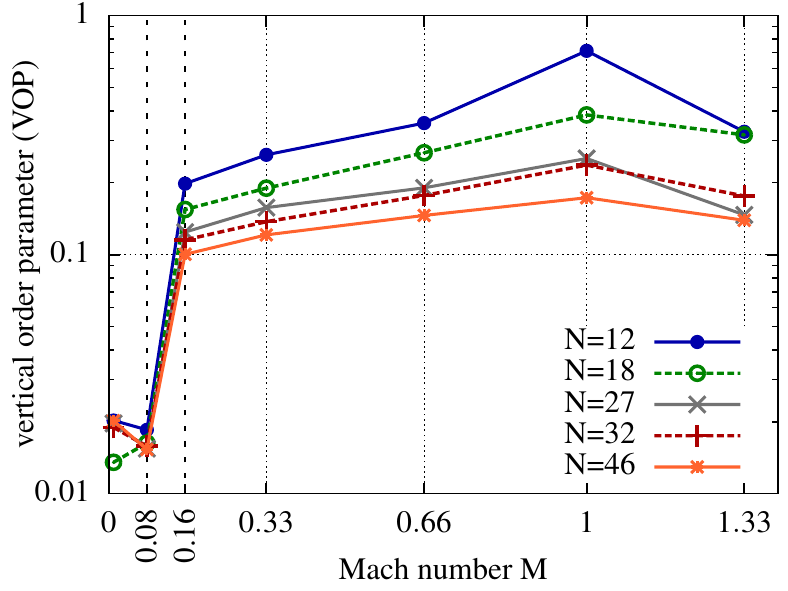}
\caption{(Color online) Vertical (string) order parameter as defined in Eq.(\ref{vop}) for different particle numbers $N$.}
\label{fig:vop}
\end{minipage}
\end{figure}

While in the two limits of $M=0$ and $M\geq1$ the grains are strongly coupled, the transition between both limits is found not to be necessarily smooth.
Instead, a relatively weak ion flow of the order of $M=0.1$ is found to completely destabilize the highly ordered dusty plasma state and the grains become spatially delocalized, see Fig~\ref{fig:dds}b). At $M=0.16$, the formation of a short-range ordered `doublet state' is observed.
The vertically aligned pairs of grains move through a weakly coupled, liquid-like state. As observed e.g. in laser manipulation experiments~\cite{melzer2001,kroll}, the downstream particle strictly follows the upper particle's motion. 
In contrast, if the lower dust particle experiences a lateral kick by other grains, the (vertical) center axis of the dust molecule is tilted sideways.
If the lower grain reaches a sufficient height with respect to the upper grain such that it leaves the attractive wakefield, the laterally repulsive grain interaction pushes the grains in the opposite directions and the binding between the grains is destroyed. 
An ion flow of $M=0.33$ is just above the threshold, where `triplet states' become stable. The lowest grain shows a strongly oscillatory behavior and other free grains or complexes are too inert to follow that motion. Moreover, the (electrostatic) interference of other grains easily destroys the binding of the lowest grain.
With further increase of $M$ more and more grains line up behind each other and form longer, streaming aligned particle chains.

A similar ion-streaming mediated order transition is found for clusters of different size. The time-averaged pair distribution function for $N_d=18$ grains in Fig.~\ref{fig:g12} gives a qualitative picture of the ion-focus induced melting transition and subsequent structure built-up upon changing $M$. 
The distinct peak structure in the two limits $M=0.01$ and $M=1.0$ is a clear signature of strong correlations in the underlying dust cluster. That is a Yukawa ball with a $(1,17)$ configuration in the static case. At ion Bohm speed, the confining potential makes it possible to have (two) nine-particles long flow-aligned strings.
These strings exhibit self-excited oscillations which grow to a large amplitude at the bottom end of the strings, as described in Ref.~\cite{lampe2005}. These ion-streaming induced oscillations are a reason for the broadening of the pair distribution at large grain distances. The pair distribution function for $M=0.08$ is (almost) structureless which indicates a fluid-like state. For $M=0.16$ we find a single sharp peak which signals the formation of temporal stable dust doublets.

To quantify the character of the string formation, we make use of the Vertical Order Parameter (VOP) which was recently introduced by Killer et al.\cite{killer}. The VOP captures the fraction of vertically aligned bonds with $\phi<10^\circ$ to the total number of bonds $B_\mathrm{tot}=N_d(N_d-1)/2$
\begin{equation} \label{vop}
 \mathrm{VOP}=\frac{B(\phi<10^\circ)}{B_\mathrm{tot}} \quad,
\end{equation}
where $\phi$ denotes the angle between the bond axis of two grains and the vertical z-axis.
The VOP is equal to one, if the particles create a single chain of particles (which may exhibit transverse oscillations \cite{schweigert}). It is zero, when there is no vertical order such as in a flat 2D system. The (time-averaged) VOP in dependence on $M$ is plotted in Fig.~\ref{fig:vop} for different particle numbers $N_d$.
As a general trend, none of the spherical clusters observed in the static case exhibits a vertical particle alignment. This finding is in agreement with Ref.~\cite{killer}. Independent of the exact particle number and finite size effects, the VOP shows a sharp jump around $c_s/10$, when the ion flow increases from $M=0.08$ to $M=0.16$. This crucial effect is due to the formation of spatially correlated doublet states. When $M$ is further increased, the formation of triplets and subsequently the formation of long particle chains is observed. Vertical alignment is  most pronounced for $M=1.0$. For $M=1.33$ the value of the VOP is slightly reduced due to a configurational rearrangement: going from $M=1.00$ to  $M=1.33$, there are for $N_d=46$ ($N_d=18$) seven (three) instead of four (two) vertical strings.

At this point, the question concerning the  mechanism for the ion-streaming induced melting process remains.
In the static case, the mutual (isotropic) grain repulsion is compensated by the external confining potential. That stabilizing interplay allows for strong correlations. At $M=0.08$, however, the grain potential in downstream direction becomes non-monotonous and relatively flat.
Consequently, the repelling force between the grains vanishes and the mechanism for the correlation built-up breaks down.
Only with further increase of the ion flow, a different kind of order process emerges which does not stabilize the system as a whole, but rather individual grains agglomerate to larger complexes (doublets, triplets, etc.) due to wakefield attraction. That is, a global stabilization process is replaced by an local one.

\section{Summary and Conclusions} \label{conclusion}
The influence of an external electric field on the dynamical alignment (laning) of like-charged particles is a topic of high current interest, e.g. in the physics of colloids~\cite{laning}.
In dusty plasmas physics, however, the important effect of the (field-induced) ion-flow on the non-equilibrium structure formation has not been systematically analysed for the full relevant range of Mach numbers $M$ so far.
To this end, we have performed first principle Langevin-type Dust Dynamics Simulations with a dynamically-screened Coulomb potential obtained from linear response theory.

The model considered here is found to reproduce the experimentally observed particle arrangements, that is Yukawa balls in the static case ($M=0$) and flow-aligned particle chains in the supersonic regime ($M\geq1$). To focus on the explicit effect of a finite Mach number $M$, the influence of $M$ on other plasma parameters such as the grain charge or trap frequency was neglected. The simulations reveal that even a relatively moderate ion-flow initiates an abrupt melting of the highly ordered (equilibrium) many-particle state. The observed, abrupt melting process resembles a non-equilibrium phase transition. 
 With further increase of the ion flow, the delocalized grains start to form flow-aligned complexes: doublets, triplets, and longer particle strings. For $M\geq1$, a re-entrance into a strongly coupled, non-equilibrium many-particle state consisting of several (oscillating) chains is observed. 
A quantitative analysis of this non-equilibrium phenomenon by means of the Vertical Order Parameter shows that the order process appears in the same $M$-range for clusters of different size. Since the underlying, wake-induced, order mechanism is of local nature, we expect a similar ion-focus induced order transition to appear in similar form also in unconfined macroscopic systems.

Since dusty plasmas are often considered as a test system for many-particle correlations, the finding that even weak streaming can trigger unexpected many-particle behavior, may be of relevance also for other fields of physics. For instance, in warm dense matter physics, electron streaming effects are often neglected in the theoretical description. Using a quantum (e.g, Mermin) dielectric function, the DDS scheme can be directly applied also to quantum systems~\cite{QDSA}.

\subsection*{Acknowledgments}

The authors thank G. Joyce and M. Lampe for sharing their in-depth knowledge about dressed particle simulations.
Furthermore, we gratefully acknowledge fruitful discussions with A. Schella, C. Killer, A. Melzer, D. Block, and K. Fujioka as well as the financial support by the DFG via grant LU 1586/1-1.

\end{document}